\documentclass[english,prl,twocolumn,showpacs]{revtex4}
\usepackage[T1]{fontenc}
\usepackage[latin9]{inputenc}
\usepackage{amsmath}
\usepackage{graphicx}
\usepackage{amssymb}

\makeatletter

\providecommand{\tabularnewline}{\\}

\@ifundefined{textcolor}{}
{%
 \definecolor{BLACK}{gray}{0}
 \definecolor{WHITE}{gray}{1}
 \definecolor{RED}{rgb}{1,0,0}
 \definecolor{GREEN}{rgb}{0,1,0}
 \definecolor{BLUE}{rgb}{0,0,1}
 \definecolor{CYAN}{cmyk}{1,0,0,0}
 \definecolor{MAGENTA}{cmyk}{0,1,0,0}
 \definecolor{YELLOW}{cmyk}{0,0,1,0}
 }

\usepackage{color}
\usepackage{bm}
\def\kbar{{\mathchar'26\mkern-9mu k}}

\makeatother

\usepackage{babel}

\begin{document}

\preprint{This line only printed with preprint option}

\title{Suppression of decoherence effects in the quantum kicked rotor}

\author{Maxence Lepers}

\altaffiliation{Present address: Laboratoire Aimé Cotton, CNRS, Universit Paris-Sud, Bt. 505, F-91405 Orsay}

\author{V\'eronique Zehnl\'e}

\author{Jean Claude Garreau}

\affiliation{Laboratoire de Physique des Lasers, Atomes et Mol\'ecules, Universit\'e
Lille 1 Sciences et Technologies; CNRS; F-59655 Villeneuve d'Ascq
Cedex}

\homepage{http://phlam.univ-lille1.fr/atfr/cq}
\begin{abstract}
We describe a method allowing transient suppression of decoherence
effects on the atom-optics realization of the kicked rotor. The system
is prepared in an initial state with a momentum distribution concentrated
in an interval much sharper than the Brillouin zone; the measure of
the momentum distribution is restricted to this interval of quasimomenta:
As most of the atoms undergoing decoherence processes fall outside
this detection range and thus are not detected, the measured signal
is effectively decoherence-free.
\end{abstract}


\maketitle
The kicked rotor has been studied for many years in the helm of both
classical and quantum Hamiltonian dynamics, where it has the status
of a paradigm. Its formal simplicity, and the richness of its dynamics
make it one of the most studied dynamical systems. Experiments started
with the observation by Raizen and co-workers \citep{Raizen:QKRFirst:PRL95}
of quantum chaos in an atom-optics realization of the kicked rotor.
An impressive number of works are devoted to different aspects of
the kicked rotor dynamics, including dynamical localization \citep{Raizen:QKRFirst:PRL95,Amman:LDynNoise:PRL98,Raizen:LDynNoise:PRL98,AP:Bicolor:PRL00,Monteiro:LocDelocDoubleKick:PRL06,AP:PetitPic:PRL06},
quantum resonances \citep{Darcy:QRes:PRL01,Fishman:KRQuantumRes:PRL02,DArcy:HighOrderQRes:PRL03,Phillips:HighOrderQuantResBEC:PRL06,AP:QuantumRes:PRA08},
sub-Fourier resonances \citep{AP:SubFourier:PRL02}, chaos-assisted
tunneling \citep{Rubins-Dunlop:ChaosAssitTunnel:Nature01,Raizen:ChaosAssistTunnel:Science01,Raizen:FluctDecChAssistTunnel:PRL02},
ratchets \citep{Monteiro:RatchetOptLatTh:PRL02,Renzoni:QuasiperiodicRatchets:PRL06,Summy:QRRarchets:PRL08},
and, recently, the observation of the Anderson metal-insulator transition
\citep{AP:Anderson:PRL08,AP:AndersonLong:PRA09,AP:PetitPic:PRL06}.
The atomic kicked rotor is a very clean and flexible system, but it
is limited by the unavoidable presence of decoherence, mainly due
to spontaneous emission \citep{Amman:LDynNoise:PRL98,Raizen:LDynNoise:PRL98}.
In the present work, we propose and analyze a method for suppressing
to a large amount, if only transiently, decoherence effects in the
kicked rotor's dynamics, which is directly applicable to existing
experimental setups.

Consider a two-level atom placed in a standing wave generated by two
counterpropagating beams of intensity $I$ and wavenumber $k_{L}$.
On the one hand, the atom may interact non-dissipatively with the
radiation by absorbing a photon in one of the beams and emitting a
photon in the \emph{other} beam by \emph{stimulated} emission. This
is a conservative elementary process in which a {}``quantum'' of
momentum $2\hbar k_{L}$ is exchanged between the atom and the field.
It generates a potential acting on the center of mass of the atom,
which turns out to be proportional to $\Omega(x)^{2}/\Delta_{L}$$\sim(I/\Delta_{L})\cos(2k_{L}x)$,
where $\Omega$ is the (resonant) Rabi frequency, $\Delta_{L}$ is
the laser-atom detuning, $(\Delta_{L}\gg\Gamma$ where $\Gamma$ is
the natural width of the excited state) \citep{CCT:Houches:90,MetcalfStratenLaserCooling}.
This {}``optical potential'' only couples momentum states separated
by multiples of $2\hbar k_{L}$, therefore, quasimomentum is conserved.
On the other hand, there can be\emph{ }spontaneous emission (SE) process
which are dissipative and thus are fatal to quantum effects. The total
momentum exchange (along the direction of the standing wave) in a
process involving SE is $\hbar k_{L}(1+\cos\theta)$ where $\theta$
is a random angle in the range $[0,2\pi)$. After such a process the
atom momentum distribution is not anymore concentrated around multiples
of $2\hbar k_{L}$.

The atom-optics realization of the kicked rotor consists in placing
laser-cooled atoms in a periodically-pulsed standing wave. The pulse
(\emph{kick}) duration $\tau$ is very short compared to the atom
dynamics, so that the pulses can be assimilated to a Dirac delta function.
The Hamiltonian for the center of mass motion of atoms of mass $M$
and momentum $p$ along the laser beam propagation direction $x$
is thus
\begin{equation}
H=\frac{P^{2}}{2}+K\cos X\sum_{m=0}^{N-1}\delta(t-m),
\label{eq:H}\end{equation}
where we introduced normalized units \citep{Raizen:QKRFirst:PRL95}
where time is measured in units of the pulse period $T$, $X=2k_{L}x$,
$P=2k_{L}Tp/M$, and $K=4\hbar k_{L}^{2}\Omega^{2}T\tau/\Delta_{L}M$.
With such definitions, $[X,P]=i\kbar$ where $\kbar=4\hbar k_{L}^{2}T/M$
plays the role of a normalized Planck constant. Note that $P/\kbar=p/2\hbar k_{L}$.

For $K\gtrsim5$ the \emph{classical} dynamics obtained from Hamiltonian~(\ref{eq:H})
is an ergodic chaotic diffusion \citep{Chirikov:ChaosClassKR:PhysRep79}
which leads to a Gaussian momentum distribution whose broadening corresponds
to an average kinetic energy increasing linearly with time, $E_{cl}=D_{cl}t$,
with $D_{cl}\sim K^{2}/4$. In the quantum case, after a characteristic
\emph{localization time} $t_{L}$, quantum interference effects lead
to a \emph{saturation} of the kinetic energy $E$ to the constant
value $E_{L}=P_{L}^{2}/2$ \citep{Casati:LocDynFirst:LNP79,Izrailev:LocDyn:PREP90},
and the average momentum distribution takes an exponential shape $\sim\exp\left(-|P|/P_{L}\right)$,
hence the name of {}``dynamical localization'' (DL) given to this
phenomenon. The instantaneous diffusion coefficient $D_{0}(t)=dE/dt$
starts with an initial value $D_{0}(0)=D_{q}$ \citep{note:Dq} and
tends to zero asymptotically. Following \citep{Amman:LDynNoise:PRL98},
we will simply model this behavior by an exponential with a characteristic
time $t_{s}$ \citep{note:ts} 
\begin{equation}
D_{0}(t)=D_{q}e^{-t/t_{s}}.
\label{eq:D0(t)}\end{equation}

DL, which is a quantum interference effect, is highly sensitive to
decoherence and will serve as a decoherence probe in the present work.
Decoherence effects in the kicked rotor have been studied both theoretically
\citep{Cohen:LocDynTheo:PRA91} and experimentally \citep{Amman:LDynNoise:PRL98,Raizen:LDynNoise:PRL98}.
Here, decoherence is essentially due to SE, whose probability per
kick is 
\begin{equation}
\Pi\equiv\frac{\Gamma\tau}{2}\frac{\Omega^{2}/2}{\Delta_{L}^{2}+\Omega^{2}/2+\Gamma^{2}/4}
\approx\frac{\Gamma\tau\Omega^{2}}{4\Delta_{L}^{2}},
\label{eq:SpEmissionPerKick}\end{equation}
where the approximate value on the right corresponds to the limit
$|\Delta_{L}|\gg\Gamma,\Omega$. Typical values of $\Pi$ in experiments
are a fraction of $10^{-2}$; DL can thus be observed only for a hundred
kicks or so \citep{AP:AndersonLong:PRA09}.

Suppose one prepares a sample of cold atoms in a well defined momentum
state, say $P=0$. Interaction with the optical potential populates
only states of momentum $P_{n}=n\kbar$, ($n$ integer) and the corresponding
wavefunction is of the form 
$\left|\psi\right\rangle =\sum_{n}c_{n}e^{i\varphi_{n}}\left|n\kbar\right\rangle $.
Stimulated processes keep \emph{well-defined} phases $\varphi_{n}$,
and allow the interference effects responsible for DL. Spontaneous
emission introduces \emph{random} phases and puts the atom into a
mixture of momentum states; in other words, it {}``resets'' quantum
interference, restoring the initial diffusion coefficient $D_{q}$.
The form of the diffusion coefficient $D(t)$ with SE is thus a weighted
mean of the evolution without SE, given by $D_{0}(t)$, and of the
effect of SE events. As has been shown in \citep{Amman:LDynNoise:PRL98},
SE restores diffusion after a time $t\thicksim t_{s}(1+\tau_{s})^{-1}$
(where $\tau_{s}\equiv\Pi t_{s}$) with an asymptotic diffusion coefficient
given by 
\begin{equation}
D_{\infty}\equiv\frac{D_{q}\tau_{s}}{1+\tau_{s}}.
\label{eq:Dinfini}\end{equation}

\begin{figure}
\begin{centering}
\includegraphics[bb=0bp 0bp 500bp 400bp,clip,width=4.5cm]{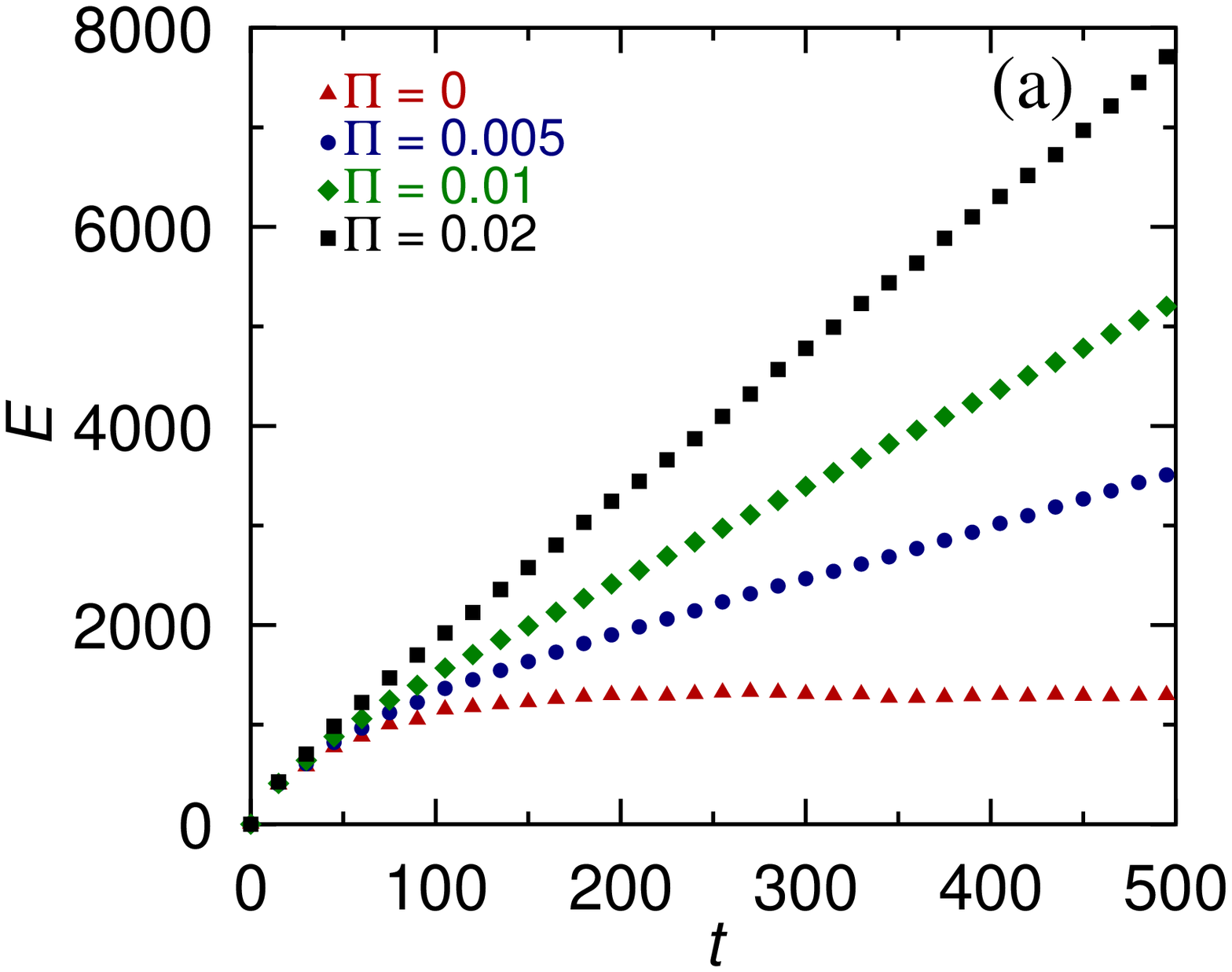}\includegraphics[width=4.5cm]{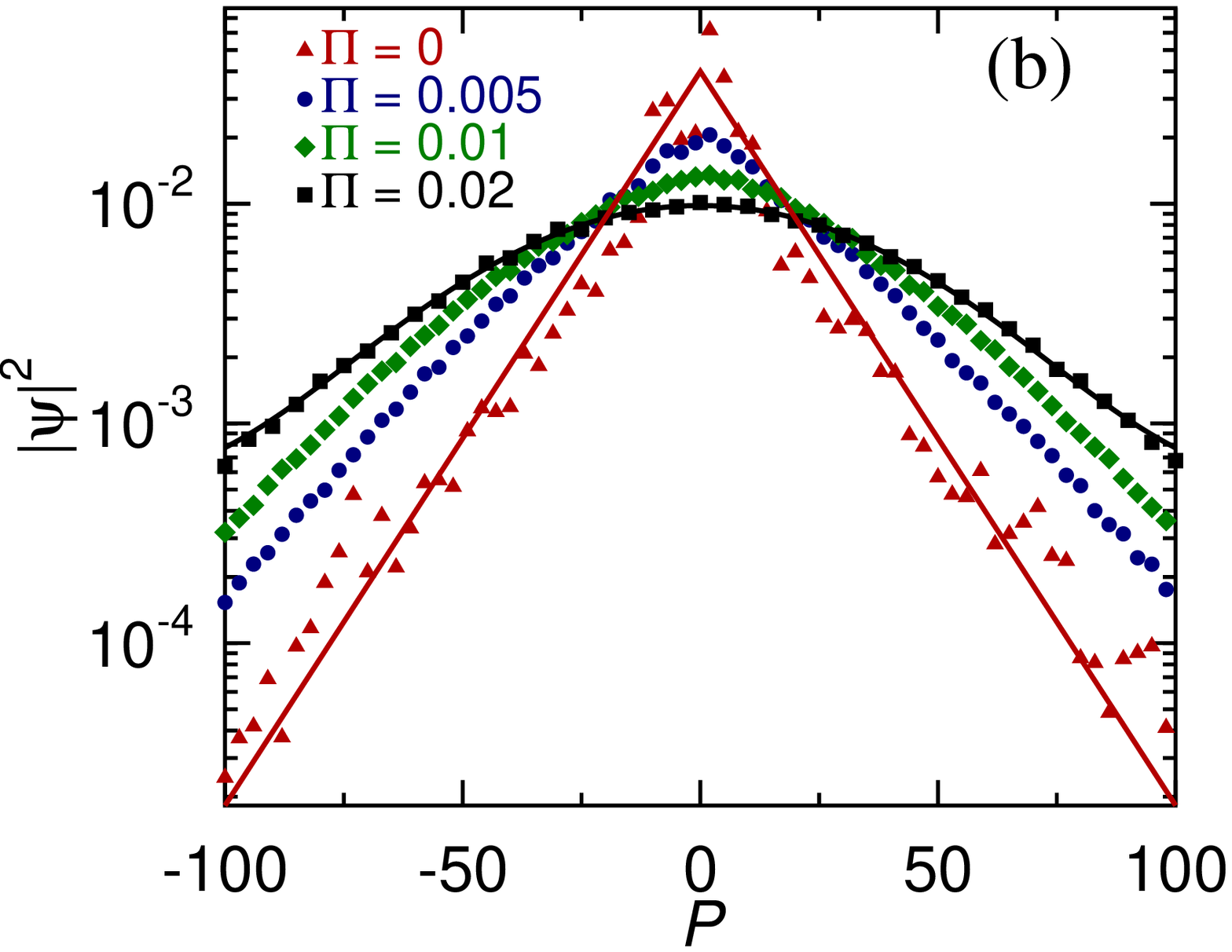}
\par\end{centering}

\caption{\label{fig:DLandDecoherence}(Color online) Effect of spontaneous
emission: (a) Kinetic energy $E(t)$ for different values of the spontaneous
emission rate $\Pi$; (b) momentum distributions (log scale) at $t=500$,
the $\Pi=0$ curve (red triangles) is well fitted by an exponential,
whereas the $\Pi=0.02$ curve (black squares) is fitted by a Gaussian.
Parameters are $K=10$, $\kbar=2.9$ ($t_{L}\approx12$) and the initial
momentum distribution is a square of width $\Delta=0.04$.}

\end{figure}

Fig.~\ref{fig:DLandDecoherence}a shows the typical behavior of the
kinetic energy for increasing levels of SE obtained by numerical simulation
of the quantum dynamics corresponding to Hamiltonian~(\ref{eq:H}).
Clearly, DL is destroyed and diffusion restored as the SE rate increases.
The initial momentum distribution is a square (for simplicity) of
width $\Delta_{R}=0.04\kbar$ centered at $P=0$. For each kick, a
Monte Carlo procedure is used to decide whether a photon is spontaneously
emitted; if so, the entire momentum distribution is translated of
a quantity $\kbar\cos\theta$, with $\theta$ picked randomly in the
interval $[0,2\pi)$, which produces the mixing of quasimomenta. In
order to simplify notations, we will measure momentum in units of
$\kbar$, so that, from now on, $P_{n}\equiv n$ and $\Delta\equiv\Delta_{R}/\kbar$.

Momentum distributions $f(P)=\left|\psi(P)\right|^{2}$, are shown
in Fig.~\ref{fig:DLandDecoherence}b. For $\Pi=0$, the distribution
is fitted by an exponential (red triangles); decoherence effects are
clearly seen for $\Pi\neq0$; for instance, the $\Pi=0.02$ is fitted
by a Gaussian (black squares). 

Let us now present our method for a transient suppression of decoherence
effects. If no SE event happens, the momentum distribution $f(P),$
at any time will be nonzero only for momenta in the range $\Delta_{n}$$=\left[n-\Delta/2,n+\Delta/2\right]$.
As $\Delta\ll1,$ atoms undergoing a SE process will mostly populate
other momentum classes. If, at the end of the kick series, one measures
the momentum distribution by \emph{probing only the $\Delta_{n}$
momentum classes}, most of the atoms having performed SE will be excluded
from the resulting signal, \emph{which will be effectively decoherence-free}.
Experimentally, preparation and measurement of $f(P)$ with a precision
$\Delta\ll1$ can be made e.g. using Raman stimulated spectroscopy
\citep{Chu:RamanCooling:PRL92,AP:RamanSpectro:PRA01,AP:Polarization:OC07}.
However, the probability that an atom having performed a SE falls
back into a detection range $\Delta_{n}$ increases with the number
of fluorescence cycles, so this filtering is a transient effect. 

\begin{figure}
\begin{centering}
\includegraphics[width=7cm]{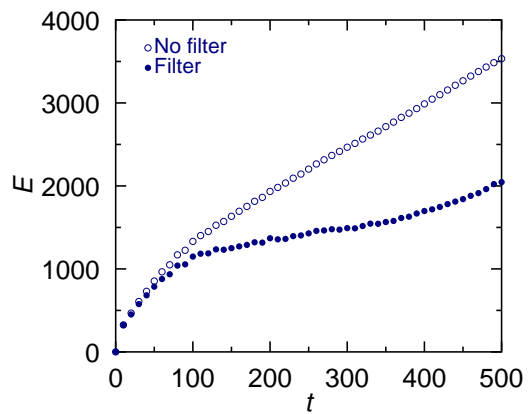}
\par\end{centering}

\caption{\label{fig:DLFiltered-1}(Color online) Average kinetic energy with
filtering (full circles) and without filtering (empty circles) for
$\Pi=0.005$ and at $N=300$ kicks (1.5 times the typical SE time,
$\Pi^{-1}=200$ kicks). The filtering reduces the average energy at
of 60 \%. Referring to the saturation value of the kinetic energy
in absence of SE, the SE effect has been reduced by a factor of about
5.5.}

\end{figure}

Fig.~\ref{fig:DLFiltered-1} shows the kinetic energy of filtered
atoms, $\bar{E}=\sum_{n}\mathcal{P}_{n}^{2}/2$ (full circles), with
$\mathcal{P}_{n}\equiv\int_{n-\Delta/2}^{n+\Delta/2}f(P)PdP$. The
comparison with the result in the case where there is no filtering
(empty circles) shows the efficiency of the filtering.

Let us make the reasonable assumption that an atom emitting spontaneously
a photon has a probability $\Delta$ of falling in one of the ranges
$\Delta_{n}$ and $(1-\Delta)$ of falling outside any of these ranges.
We then consider three populations of atoms: $F_{0}$, the population
of atoms which have \emph{never} performed (at time $t$) any SE (the
momentum of these atoms is thus in the ranges $\Delta_{n}$); $F_{\Delta}$,
the population of atoms in one of the ranges $\Delta_{n}$ but having
performed at least one SE, and $F_{1-\Delta}$, the population of
atoms outside any detection range $\Delta_{n}$. The rate equations
for these populations are straightforwardly obtained (noting that
$F_{0}+F_{\Delta}+F_{1-\Delta}=1$) as:\begin{eqnarray}
\frac{dF_{0}}{dt} & = & -\Pi F_{0}\label{eq:PopulationRateEqs0}\\
\frac{dF_{\Delta}}{dt} & = & -\Pi F_{\Delta}+\Pi\Delta,\label{eq:PopulationRateEqsDelta}\end{eqnarray}
 The integration of these equations gives:\begin{eqnarray}
F_{0}(t) & = & e^{-\Pi t}\label{eq:F0(t)}\\
F_{\Delta}(t) & = & \Delta\left(1-e^{-\Pi t}\right).\label{eq:FDelta(t)}\end{eqnarray}
Note that the population of \emph{detected} (or filtered) atoms $F_{0}+F_{\Delta}$$=\Delta+(1-\Delta)e^{-\Pi t}$
is not constant in time and tends, as expected, to $\Delta$ as $t\rightarrow\infty$.

The kinetic energy of \emph{filtered} atoms is 
\begin{equation}
\overline{E}(t)=\frac{E_{0}+E_{\Delta}}{F_{0}+F_{\Delta}},
\label{eq:Efiltre}\end{equation}
where $E_{0}$ and $E_{\Delta}$ are the \emph{total} kinetic energy
calculated over, resp., the populations $F_{0}$ and $F_{\Delta}$.
The evolution of $E_{0}$ can be obtained from the equation $dE_{0}/dt=D_{0}(t)F_{0}-\Pi E_{0}$:
the first term is the contribution due to the kicks {[}see Eq.~(\ref{eq:D0(t)}){]}
and the second term is the depletion by SE. Thus
\begin{equation}
E_{0}(t)=D_{q}t_{s}\left(1-e^{-t/t_{s}}\right)e^{-\Pi t},
\label{eq:E0(t)}\end{equation}
which has a maximum at time 
\begin{equation}
t_{1}=t_{s}\ln(1+\tau_{s}^{-1}).
\label{eq:t1}\end{equation}

The evolution of $E_{\Delta}$ is governed by two contributions: \emph{i})
the contribution of the atoms in the population $F_{\Delta}$ which
performed at least one SE event and whose kinetic energy is due to
all past fluorescence cycles \citep{Cohen:LocDynTheo:PRA91}, and
\emph{ii}) the contribution of the atoms in the population $F_{0}$
that perform a fluorescence cycle at time $t$ and fall into $F_{\Delta}$,
that is 
\begin{equation}
\frac{dE_{\Delta}}{dt}=\int dt^{\prime}\frac{dF_{\Delta}}{dt^{\prime}}D(t^{\prime})+\Delta\Pi E_{0}(t),
\end{equation}
which gives, after some algebra:
\begin{align}
E_{\Delta} & =\frac{\Delta D_{q}t_{s}}{1+\tau_{s}}\left(\Pi t-(1+\tau_{s})e^{-\Pi t}\right.\nonumber \\
 & \left.+\frac{\tau_{s}(1+2\tau_{s})}{1+\tau_{s}}e^{-(1+\tau_{s})t/t_{s}}+
\frac{1+\tau_{s}-\tau_{s}^{2}}{1+\tau_{s}}\right).
\label{eq:Edelta(t)}\end{align}
From Eqs.~(\ref{eq:Efiltre}), (\ref{eq:E0(t)}) and (\ref{eq:Edelta(t)})
one obtains an explicit expression for $\overline{E}(t)$ that we
do not give here as it somewhat cumbersome; let us just consider the
limit $\Delta,\tau_{s}\ll1$ in which it can be written:
\begin{equation}
\bar{E}(t)=D_{q}\tau_{s}\Pi\Delta e^{\Pi t}\left(1+\Pi t\right).
\label{eq:E(t)approx}\end{equation}
\begin{figure}
\begin{centering}
\includegraphics[width=4.5cm]{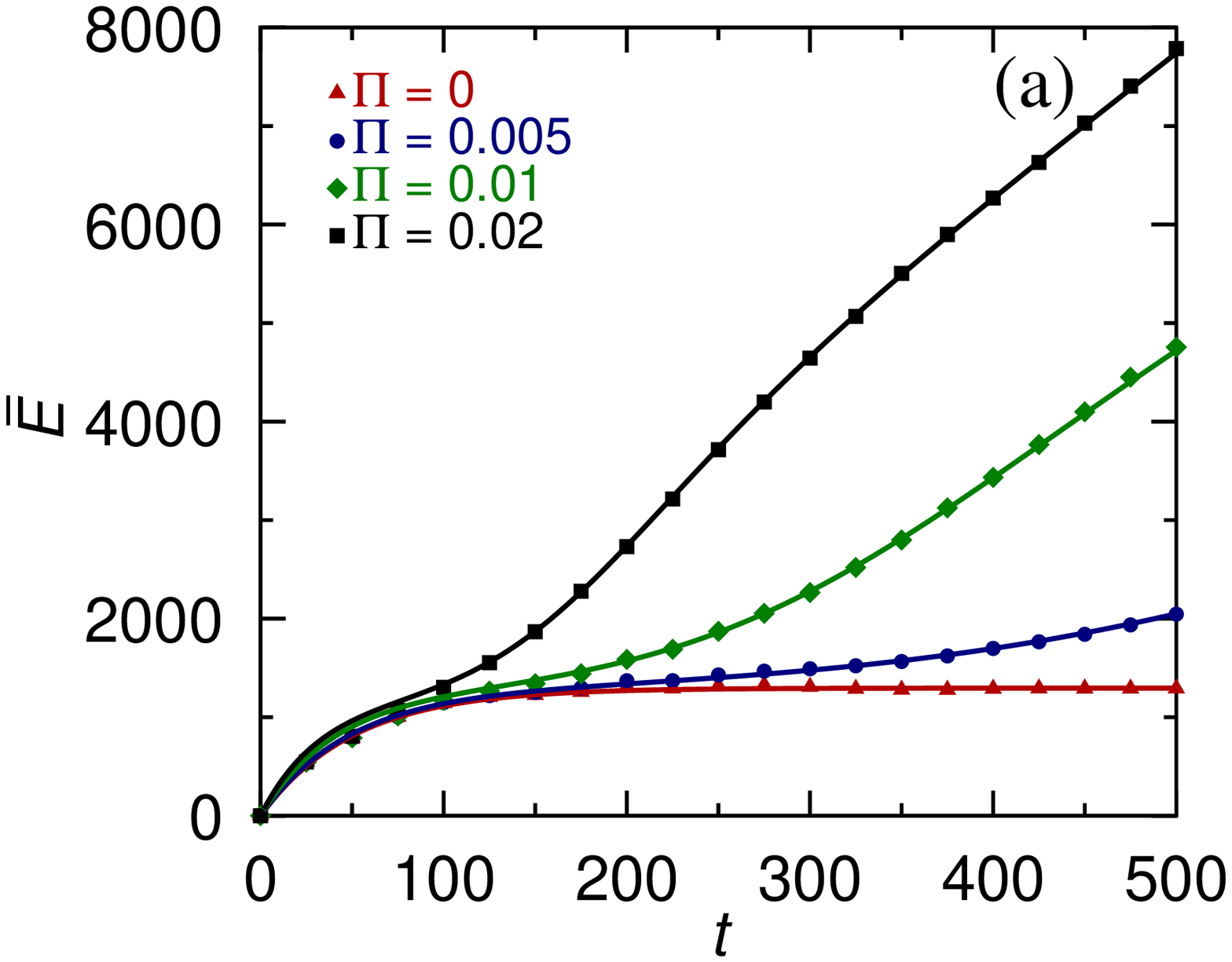}\includegraphics[width=4.5cm]{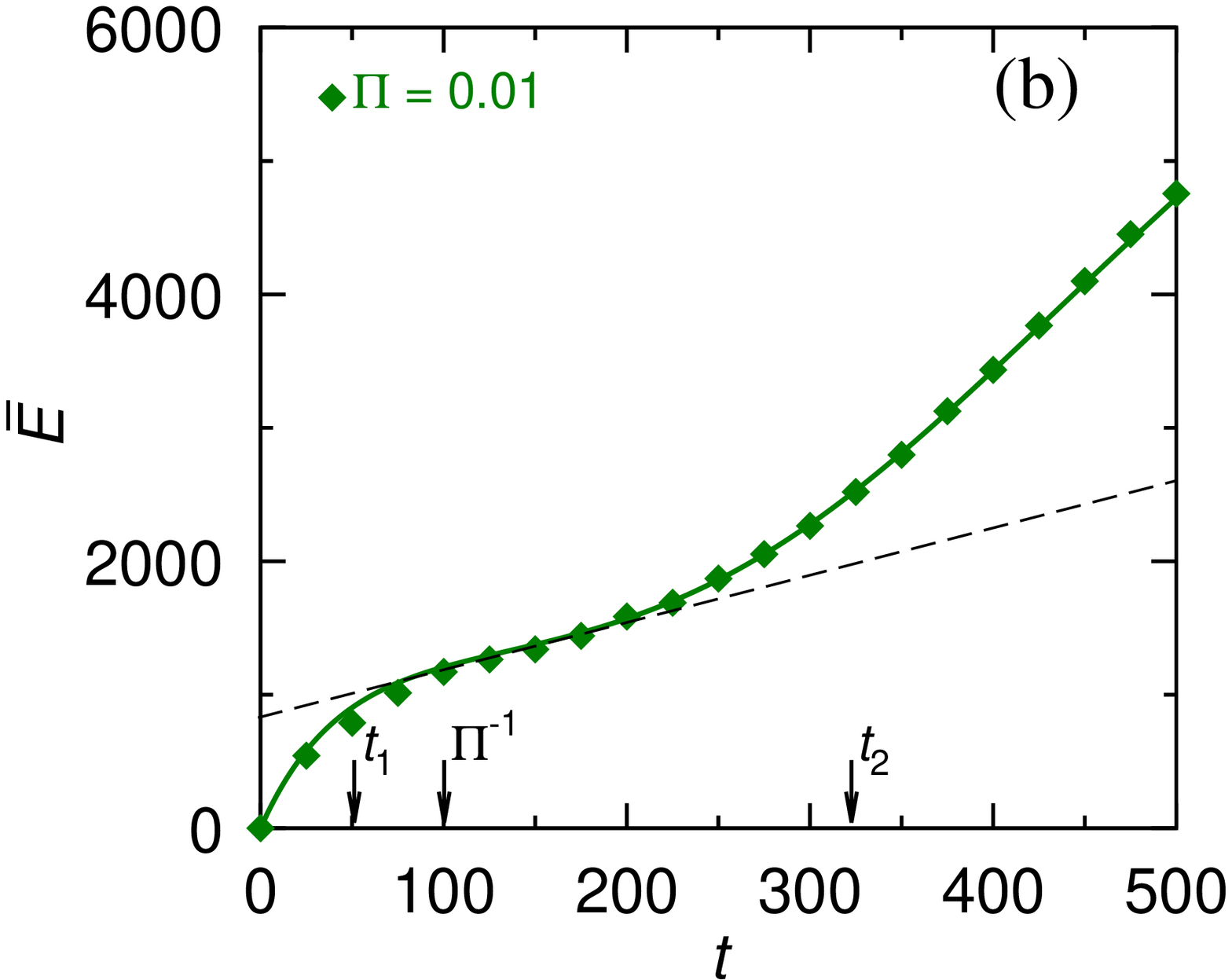}
\par\end{centering}

\caption{\label{fig:DLFiltered-2}(Color online) (a) Plot of $\overline{E}(t)$
obtained from the numerical simulation (same curves as in Fig.~\ref{fig:DLandDecoherence})
and best fits obtained from Eq.~(\ref{eq:Efiltre}) with the fit
parameters $D_{q}$ and $t_{s}$ (full lines). (b) The $\Pi=0.01$
curve of plot (a), indicating the characteristic times $t_{1}\approx51$
and $t_{2}\approx320$, and the reduced diffusion (dashed line) with
coefficient $D_{r}$ (cf. text).}

\end{figure}
Fig.~\ref{fig:DLFiltered-2}a shows the numerical results for $\overline{E}(t)$
for different values of $\Pi$. The full lines are fits of the numerical
data obtained from Eqs.~(\ref{eq:Efiltre}), (\ref{eq:E0(t)}) and
(\ref{eq:Edelta(t)}), using $D_{q}$ and $t_{s}$ as adjustable parameters
(see notes \citep{note:Dq,note:ts}), whose values are given in Table
\ref{tab:FitParam}.

\begin{table}
\begin{centering}
\begin{tabular}{|c|c|c|c|c|}
\hline 
$\Pi$ & 0 & 0.005 & 0.01 & 0.02\tabularnewline
\hline
\hline 
$D_{q}$ & 25.5 & 26.3 & 30.7 & 36.6\tabularnewline
\hline 
$t_{s}$ & 50.7 & 48.8 & 41.3 & 32.5\tabularnewline
\hline
\end{tabular}
\par\end{centering}

\caption{\label{tab:FitParam}Fit parameters $D_{q}$ and $t_{s}$ corresponding
to the curves displayed on Fig.~\ref{fig:DLFiltered-2}.}

\end{table}

The main features of the curves in Fig.~\ref{fig:DLFiltered-2} can
be understood by simple arguments. The curve in Fig.~\ref{fig:DLFiltered-2}b
shows two characteristic times: \emph{i}) time $t_{1}$ {[}Eq.~(\ref{eq:t1}){]}
at which the SE starts depleting the population $F_{0}$, \emph{ii})
time $t_{2}$ when the contribution of the {}``incoherent atoms''
$F_{\Delta}$ exceeds the {}``coherent'' contribution due to $F_{0}$;
$t_{2}$ is thus given by the condition $F_{\triangle}(t_{2})=F_{0}(t_{2})$
which leads to $t_{2}\approx-\ln\Delta\Pi^{-1}$. In the absence of
filtering, $t_{2}\sim\Pi^{-1}$, the filtering thus increases this
time by a factor $|\ln\Delta|$.

The energy $\overline{E}(t)$ exhibits a transient reduced diffusion
regime that corresponds to the quasi-plateau which is clearly seen
in Fig.~\ref{fig:DLFiltered-2}b; the corresponding diffusion coefficient
$D_{r}$ (indicated by the dashed line) can be estimated by expanding
Eq.~(\ref{eq:E(t)approx}) around $t=\Pi^{-1}$, which gives $D_{r}\approx2eD_{q}\tau_{s}\Delta$$\approx2e(1+\tau_{s})\Delta D_{\infty}$
with $D_{\infty}$ given by Eq.~(\ref{eq:Dinfini}). Thus the condition
for the efficiency of filtering is $\Delta<\left[2e(1+\tau_{s})\right]^{-1}$($\sim0.1$
for current parameters).

\begin{figure}
\begin{centering}
\includegraphics[width=7cm]{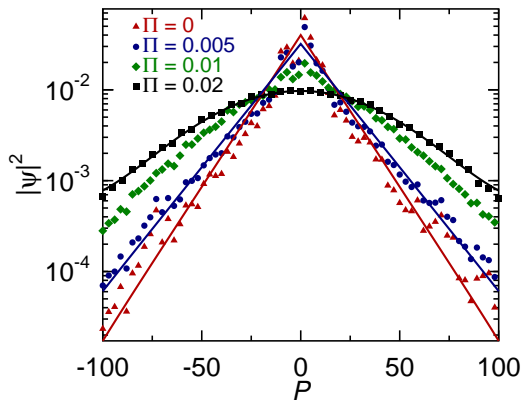}
\par\end{centering}

\caption{\label{fig:FilteredMomentumDistrs}(Color online) Momentum distributions
at $N=500$. The $\Pi=0$ (red triangles) and $\Pi=0.005$ (blue circles)
curves are fitted by exponentials, while the $\Pi=0.02$ (black squares)
curve is well fitted by a Gaussian. The curve corresponding to $\Pi=0.01$
(green diamonds) has an intermediate shape.}

\end{figure}

Finally, let us consider the effect of filtering process on momentum
distribution. Fig.~\ref{fig:FilteredMomentumDistrs} displays momentum
distributions after $N=500$ kicks, corresponding to the curves in
Fig.~\ref{fig:DLFiltered-2}a. One sees that in the case $\Pi=0.005$
($\Pi N=2.5)$ the distribution can still be fitted by an exponential,
showing that DL has survived up to 500 kicks (cf. the corresponding
kinetic energy curve in Fig.~\ref{fig:DLandDecoherence}). The $\Pi=0.02$
($\Pi N=10$) curve is fitted by a Gaussian and is thus affected by
decoherence. The case $\Pi=0.01$ ($\Pi N=5)$ has an intermediate
shape between exponential and Gaussian. 

In conclusion, we have proposed an original method allowing the suppression
of decoherence effects in the dynamics of the quantum kicked rotor
for a finite but quite long time compared to the typical duration
of current experiments. The method can be applied to state of art
experiments and shall allow more precise studies of systems in which
decoherence is the main limitation \citep{AP:Anderson:PRL08,AP:AndersonLong:PRA09}.
Although spontaneous emission was the only source of decoherence considered
here, it is worth noting that the the method is in principle applicable
to any source of decoherence which does not conserve quasimomentum
(e.g. collisions).
\begin{acknowledgments}
The authors are grateful to D. Delande for fruitful discussions.

\end{acknowledgments}

\begin{thebibliography}{31}
\bibitem[{not({\natexlab{b}})}]{note:ts}
\bibinfo{note}{{The time $t_s$ is related, but not identical, to the
  localization time $t_L$. The exact form of the saturating dynamical
  localization curve is not exponential, as it can be seen from numerical
  simulations, but this simple choice allows us reproduce the essential
  characteristics of the phenomenon. We shall consider in the following $t_s$
  as a fitting parameter.}}
\end{thebibliography}

\end{document}